\documentclass[10pt]{article}
\usepackage{url}
\newcommand{\noi}{\noindent}
\usepackage{perpage}
\MakePerPage{footnote}

\begin{document}
\bibliographystyle{osajnl}

\title{What is a photon?}
\author{Vasant Natarajan\thanks{email: vasant@physics.iisc.ernet.in} \\
{\small \em Department of Physics, Indian Institute of Science, Bangalore 560 012, India}}

\date{\today}

\maketitle

\begin{abstract}
We discuss the absorber theory of radiation as put forward by Wheeler and Feynman. We show that it gives a better understanding of the photon compared to the usual quantum electrodynamics (QED) picture.

\vspace{2mm}


\end{abstract}

\noi
{\em All the fifty years of conscious brooding have brought me no closer to answer the question, `What are light quanta?' Of course today every rascal thinks he knows the answer, but he is deluding himself.} \\
\mbox{} \hfill --- Albert Einstein

\vspace{10mm}

Light is a propagating disturbance of the electromagnetic field. It appears as the solution of a wave equation resulting from the four Maxwell's equations in source-free region. Not surprisingly, it was treated as a classical wave, and seemed to have all the properties that one associates with a wave---interference, diffraction, reflection and refraction,  coherence, etc. Then came the mystery of blackbody radiation spectrum, which was inexplicable from this classical wave picture. In a stroke of genius, Max Planck (in 1900) made the {\em ad hoc} proposal that the energy of the emitted radiaton is {\em quantized} in units of the frequency ($E=h\nu$), and with this assumption, he could explain all the features of the spectrum. This ushered in the ``quantum era'', and caused, in the terminology of Thomas Kuhn, a paradigm shift in our understanding of nature. But the quantization of light was only an implicit idea in Planck's theory. The explicit nature of the light quantum, or photon as it is called now, came with its use by the {\em young} Einstein (in 1905) in explaining the {\em photoelectric effect}. He went on to win the Nobel Prize for this work, because this explanation firmed up the photon concept in the thinking of scientists, and the (additional) particle nature of light came to be accepted. Things came a full circle when de Broglie (in 1924) introduced the idea of wave nature for particles of matter, showing that {\em wave-particle duality} is a fundamental property of everything in nature, matter and its interactions.

A century later, most of us know how to work with photons. The advent of {\em lasers}, a coherent source of photons, has put an indispensable tool in the hands of scientists and engineers. Lasers are used everywhere today---in your computer hard drive, in bar-code scanners in shops, in laser pointers, in DVD players, in all kinds of surgery including delicate surgery of the eye, in metal cutting, in the modern research laboratory, to name a few. We, in our atomic physics laboratory, also use lasers all the time. We use them for laser cooling, to cool atoms down to a temperature of a millionth of a degree above absolute zero. We use lasers as optical tweezers, to trap micron-sized beads and cells. We use lasers in high-resolution spectroscopy, to understand the structure of atoms and validate fundamental theories.

In short, we know how to use photons, and how to use them well. But do we understand them? Perhaps not. We certainly have a useful mental picture of a straight-line beam of particles traveling at the speed of light $c$. In fact, we believe that we can actually see a laser beam---think of the familiar red line coming out of a laser pointer. But a moment's introspection will make us realize that what we are ``seeing'' is actually those photons that scatter into our eye from the ever-present dust particles in the room. Indeed, it is quite illuminating (pun intended) to see a light beam entering a vacuum chamber through a window---it seems to disappear after the window because there are no particles inside the vacuum chamber to scatter the light. A simpler experiment can be done if you have access to a plane polarized beam. If the polarization axis is oriented in the vertical direction, then you will not see the beam if you view it from the top. This is because the scattering probability is exactly {\em zero} along the polarization axis. Therefore, when we say we ``see'' something, what we are talking about is that some photons have reached our retina.

Consider the phenomenon of spontaneous emission. One learns that an atom in an excited state ``wants'' to go to the lowest-energy ground state. What do you mean ``wants''? Atoms do not have feelings. The excited state is as good a solution of the Hamiltonian of the atom as the ground state. Every state is a stable {\em stationary} solution (called an eigenstate), just that they all have different energies. So why is a lower energy better, and the lowest energy the best? What, in fact, {\em causes} spontaneous emission, i.e.\ {\em induces} the atom to go from a higher to a lower state? Our latest and most-successful theory to date---quantum electrodynamics (QED, see Box 1)---says that spontaneous emission is actually stimulated emission, but one where the stimulation is from the vacuum modes. This may be a clever way of doing calculations, but it is unsatisfactory because the total energy in the vacuum modes (called the zero-point energy) is {\em infinity}. This is one of several infinities that plague QED; we know how to work around these infinities, but it still leaves a bad taste in the mouth.

Equally puzzling is the phenomenon of {\em photon recoil}, also known as {\em radiation reaction}. This is the momentum kick that an atom\footnote{I use the word ``atom'' to mean any piece of matter.} receives when it emits a photon, similar to the recoil that you feel when you fire a gun. The bullet is a real particle that carries momentum, and the recoil kick is just a consequence of momentum conservation. But the photon recoil is due to the momentum transferred by a massless particle of {\em interaction}. This recoil effect is real, in fact Einstein used it in his 1917 paper to predict the phenomenon of stimulated emission \cite{NAT01reso}. And the same momentum transfer is used for the well-known phenomenon of laser cooling. But, unless the photon is given {\em independent} reality, the mechanism by which the momentum of the atom changes cannot be understood.

Which brings us to the question---is the photon independently real? Let us not forget that light is an interaction between electrical charges. The big-bang model of cosmology says that there was a time in the early universe when only photons were present. It seems illogical to say that the early universe was full of interactions, but had no matter between which the interactions could occur. It is like saying there is a room full of conversations, but no people to converse between. Conversation is an interaction between people. No people, no conversation.

Enter Wheeler and Feynman, and their paper titled ``Interaction with the Absorber as the Mechanism of Radiation'' \cite{WHF45}. They show that the photon is {\em not} independently real, and give a satisfactory answer to all of the above puzzles. In fact, the puzzles---especially that of radiation reaction---were known for a long time, and many scientists (like Fokker and Schwarzchild) had proposed solutions. The idea that Wheeler and Feynman developed was based on an earlier proposal by Tetrode \cite{TET22}, a fact that was pointed out to them by Einstein. As they write in a footnote in the paper: \\

\footnotesize
\noi
When we gave a preliminary account of the considerations which appear in this paper (Cambridge meeting of the American Physical Society, February 21, 1941, Phys.\ Rev.\ {\bf 59}, 683 (1941)) we had not seen Tetrode's paper.  We are indebted to Professor Einstein for bringing to our attention the ideas of Tetrode and also of Ritz, who is cited in this article.  An idea similar to that of Tetrode was subsequently proposed by  G.\ N.\ Lewis, Nat.\ Acad.\ Sci.\ Proc.\ {\bf 12}, 22 (1926): ``I am going to make the \ldots assumption that an atom never emits light except to another atom, and to claim that it is as absurd to think of light emitted by one atom regardless of the existence of a receiving atom as it would be to think of an atom absorbing light without the existence of light to be absorbed.\footnote{I would add that it is equally absurd to think of a universe with only light and no atoms to emit or absorb it, {\em apropos} my previous comment.} I propose to eliminate the idea of mere emission of light and substitute the idea of {\em transmission}, or a process of exchange of energy between two definite atoms or molecules.'' Lewis went nearly as far as it is possible to go without explicitly recognizing the importance of other absorbing matter in the system, a point touched upon by Tetrode, and shown below to be essential for the existence of the normal radiative mechanism. \\

\normalsize
The idea of Tetrode also is to abandon the concept of electromagnetic radiation as an elementary process and to interpret it as a consequence of an {\em interaction} between a source and an absorber. His exact words are worth repeating: \\

\footnotesize
\noi
The sun would not radiate if it were alone in space and no other bodies could absorb its radiation \ldots. If for example I observed through my telescope yesterday evening that star which let us say is 100 light years away, then not only did I know that the light which it allowed to reach my eye was emitted 100 years ago, but also the star or individual atoms of it knew already 100 years ago that I, who then did not even exist, would view it yesterday evening at such and such a time. One might accordingly adopt the opinion that the amount of material in the universe determines the rate of emission.  Still this is not necessarily so, for two competing absorption centers will not collaborate but will presumably interfere with each other. If only the amount of matter is great enough and is distributed to some extent in all directions, further additions to it may well be without influence. \\

\normalsize
Radiation reaction was well known from the fact that a charged particle on being accelerated loses energy by emitting radiation. This loss can be interpreted as being caused by a
force acting on the particle given in magnitude
and direction by the expression
\[
\frac{2({\rm charge})^2({\rm time \; rate \; of \; change \; of \; acceleration})}{3( {\rm velocity \; of \; light})^3}
\]
when the particle is moving slowly. Wheeler and Feynman take up the proposal of Tetrode to get two results: the above expression for radiation reaction, and that the fields we are familiar with from experience are all time retarded. For this, they give his idea the following definite formulation:

\footnotesize

\begin{enumerate}
\item An accelerated point charge in otherwise charge-free space does not radiate electromagnetic energy.

\item The fields which act on a given particle arise only from other particles.

\item These fields are represented by one-half the retarded plus one-half the advanced Lienard-Wiechert solutions of Maxwell's equations. This law of force is symmetric with respect to past and future.\footnote{We now have some evidence that the fundamental laws of physics violate such time-reversal symmetry. One consequence of this would be the existence of a permanent electric dipole moment (EDM) in an atom or molecule, though none has been found so far. Therefore, EDM searches, which is also being done in my lab, are among the most-important experiments in physics today.} In connection with this assumption we may recall an inconclusive but illuminating discussion carried on by Ritz and Einstein in 1909, in which Ritz treats the limitation to retarded potentials as one of the foundations of the second law of thermodynamics, while Einstein believes that the irreversibility of radiation depends exclusively on considerations of probability.  Tetrode, himself, like Ritz, was willing to assume elementary interactions which were not symmetric in time. However, complete reversibility is assumed here because it is an essential element in a unified theory of action at a distance.  In proceeding on the basis of this symmetrical law of interaction, we shall be testing not only Tetrode's idea of absorber reaction, but also Einstein's view that the one-sidedness of the force of radiative reaction is a purely statistical phenomenon.  This point leads to our final assumption:

\item Sufficiently many particles are present to absorb completely the radiation given off by the source.

\end{enumerate}

\normalsize
\noi
As mentioned in point 3, this is a theory of action at a distance, but not the kind of {\em instantaneous} action at a distance envisaged by Newton for his theory of gravitation. It is action propagated at a finite velocity, in this case the velocity of light.

In this picture, the absorber is the {\em cause} of radiation. When the absorber receives the photon, it {\em moves}, or more correctly accelerates. Therefore, processes such as spontaneous emission and radiation reaction are caused by the advanced field of this movement appearing at the source. The half-advanced field is essential so that this cause appears at the exact instant of radiation---the recoil felt by the atom is simultaneous with the emission of the photon. If the retarded and advanced fields due to acceleration of the source are $F_{\rm ret}$ and $F_{\rm adv}$ respectively, then the total field emanating from the source is
\[
\frac{F_{\rm ret}}{2} + \frac{F_{\rm adv}}{2} \, .
\]
Wheeler and Feynman show that the total field near the source due to all the absorbers is
\[
\frac{F_{\rm ret}}{2} - \frac{F_{\rm adv}}{2} \, .
\]
This field was called the ``radiation field'' by Dirac, and its form was {\em assumed} by him in order to get the correct expression for the radiation reaction. Now, we have an explanation for its origin. Moreover, the complete field diverging from the source that would be felt by a test particle (which is just the sum of the above two terms), is {\em the full retarded field}, as required by experience.

We see that the above picture gives a self-consistent explanation of radiation. To quote from the paper: \\

\footnotesize
\noi
Our picture of the mechanism of radiation is seen to be self-consistent. Any particle on being accelerated generates a field which is half-advanced and half-retarded. From the source a disturbance travels outward into the surrounding absorbing medium and sets into motion all the constituent particles. They generate a field which is equal to half the retarded minus half the advanced field of the source. In this field we have the explanation of the radiation field assumed by Dirac. The radiation field combines with the field of the source itself to produce the usual retarded effects which we expect from observation, and such retarded effects only. The radiation field also acts on the source itself to produce the force
of radiative reaction. What we have said of one particle holds for every particle in a completely absorbing medium. All advanced fields are concealed by interference. Their effects show up directly only in the force of radiative reaction. Otherwise we appear to have a system of particles acting on each other via purely retarded forces. \\

\normalsize

Wheeler and Feynman next show that the irreversibility of radiation is not due to electrodynamics itself but due to the statistical nature of absorption, {\em \`a la} Einstein. To understand this, it is enlightening to compare radiation with heat conduction. Both processes convert ordered into disordered motion although every elementary interaction involved is microscopically reversible. In heat conduction, an initially hot body cools off with time because the {\em probability} for cooling is overwhelmingly greater than the chance for it to grow hotter. Similarly, if we start with a charged particle whose energy is large in comparison to the surrounding absorber particles, then there is an overwhelming probability that the particle will lose energy to the absorber (at a rate in close accord with the law of radiative damping). Take the classic example of the irreversible breaking of an egg. If we could choose the initial conditions so that the millions of particles involved had exactly the reverse of the motion acquired during breakage, we would see an egg forming from its constituent pieces. It is just that the probability of this happening is negligible.

The expression for the force of radiation reaction shows that it is proportional to the first derivative of acceleration, or the third derivative of position. This means that a charge starts to move before the arrival of the disturbance; and $e^2/mc^3$ seconds ahead of the time when it attains a velocity comparable with its final speed. This has been termed {\em pre-acceleration}. Since the disturbance in this case is the advanced field of the absorber, we have to give up the notion that the movement of a particle at a given instant is completely determined by the motions of all other particles at earlier moments. Pre-acceleration can be hence viewed as an influence of the future on the past, i.e.\ the distinction between past and future is blurred on time scales of the order of $e^2/mc^3$. In other words, {\em those phenomena which take place in times shorter than this figure require us to recognize the complete interdependence of past and future in nature, an interdependence due to an elementary law of interaction
between particles which is perfectly symmetrical between advanced and retarded fields}.

The absorber picture of radiation seems ``repugnant to our notions of causality'' \cite{LEW26}, in the sense that we can (at least in principle) change the process of emission by intervening suitably---by blocking the path from emitter to absorber for example. Without bringing notions of human free will and philosophical complications involving life, let us imagine a simple intervention scheme where a shutter is designed to (automatically) block the path of the photon halfway between the source and absorber. Does the photon go back to the source and re-excite it because the path to the absorber is now blocked? No. The correct solution which comes out of the absorber theory is that the {\em advanced} field of the shutter tells the atom not to radiate in the first place. That is why the advanced field is so important to this theory, it gives a consistent solution irrespective of the distance between the absorber and emitter---your eye and the light from a distant star millions of light years away, for example.

The theory also gives a satisfactory explanation for the well-known phenomenon of {\em photonic bandgap}. This is a system where a periodic array of dielectric materials is used to create a bandgap for light---a situation where the system does not allow the propagation of light waves with certain energies or wavelengths. This is akin to the bandgap for electrons in a crystal, where the periodic array of nuclei creates a (Bragg-scattering) condition so that certain electron waves cannot propagate. One can therefore suppress spontaneous emission from an atom in the excited state by placing it within a photonic-bandgap material, with a band gap in the correct range. This is easily understood in the absorber picture as arising due to the fact that the field of the absorber is not allowed to reach the atom. In the conventional picture, this is explained by saying the bandgap material creates a ``better vacuum'', one where the vacuum modes are suppressed.

The above analysis shows that the process of emission is {\em nonlocal}---the states of the emitter and absorber are coupled no matter how far apart they are. Nonlocality is an inherent part of quantum mechanics, and John Bell showed that it can be experimentally tested using what are now called Bell's inequalities. Most of these tests are done with entangled photons. And experimenters try to enforce locality by changing the (polarization) state of the detector while the photons are in flight. But the absorber picture tells us that the emitting atom ``knows'' the final state of the detector in advance. So there is no possibility of a ``delayed choice''---a phrase coined by Wheeler to indicate changes made after a particle has chosen one of two paths in an interference experiment. The correct way to test quantum nonlocality is to use entangled pieces of matter (two atoms dissociating from a paired singlet state, for example), and not photons.

The one puzzle that remains in the above picture is the phenomenon of {\em pair creation}, a process where a photon of suitable energy gets converted into matter consisting of a particle and antiparticle pair. Here, suitable energy means that it is at least equal to $2mc^2$, the total rest mass energy of the pair. The theory of relativity tells us that matter and energy are equivalent: the famous relation $E=mc^2$, which gives the above requirement for the minimum photon energy for pair production. But to say that a photon (remember that it is a particle of interaction) can be converted into matter seems absurd. Or to say that matter can be created out of pure energy. A better solution is that the particle-antiparticle pair was always there, but in a bound state that did not interact with other matter (or was ``invisible''). This state then absorbed a photon and disassociated into its constituent pair. No matter is ever created or destroyed, just that a photon takes the matter from being invisible to visible. Of course, this picture is valid only if such a non-interacting state is shown to exist. Note that such a state is different from the ground state of the well-known {\em positronium} atom formed using an electron and a positron (anti-electron), which is like a hydrogen atom but with the proton replaced by the positron. The positronium atom can absorb photons of much smaller energy because it has excited states that are analogous to the excited states of hydrogen.

I met Wheeler at a conference in honour of his 80th birthday at the University of Maryland in 1994. I asked him why the seemingly good theory of absorber interaction was not widely accepted. He said something (which I confess I did not fully understand) about empirically looking for complete absorption in all directions in the sky, and not finding it. In the paper, Wheeler and Feynman do discuss the consequences of incomplete absorption. The problems depend to a large extent on the model of the universe, and the description of electromagnetism in curved spacetime. Anyway, I think (in agreement with Lewis) that emission without absorption is not possible, so there is no question of partial absorption. Certainly, the present model of the photon and radiation has many puzzling features that make it unsatisfactory. To paraphrase Einstein, perhaps {\em we are deluding ourselves into thinking that we know the photon}.


\section*{Box 1}
Quantum electrodynamics (QED for short), despite its shortcomings, is arguably the pinnacle of any quantum theory to date. The same Feynman, who is the coauthor of the article under discussion, was instrumental in developing the theory. He later shared the Nobel prize for his work in QED, and called it {\em the strange theory of light and matter}. Strange indeed, but also immensely successful. In fact, it can be called our most successful theory since its prediction of the anomalous magnetic moment, or $(g-2)$, of the electron has been verified to an unprecedented accuracy of 12 digits! Its success means that it has managed to capture some inherent description of the workings of nature, so that any future theory has to at least reproduce its quantitative results. Indeed, it is our best example of a quantum field theory, with its naturally occurring creation and annihilation operators. It is used as a model for formulating field theories for other interactions---a {\em canonical} field theory if ever there was one.

The idea of my article is to show that there are alternate, {\em more-understandable} ways of looking at light. Feynman could not quantize the absorber theory in a satisfactory manner so as to get the experimental results of the $(g-2)$ of the electron and the Lamb shift of the hydrogen atom. But a reading of Narlikar's article in this issue shows that there might be ways of doing it. In the immortal words of Wheeler:
\begin{list}{}{}
\item {\em Behind it all is surely an idea so simple, so beautiful, that when we grasp it---in a decade, a century, or a millennium---we will all say to each other, how could it be have been otherwise?  How could we have been so stupid?}
\end{list}
I too think that we will find such a beautiful idea to explain the puzzles of light, simple enough that it can be explained to high-school students.

\section*{Suggestions for further reading}
\begin{enumerate}
\item J.~A. Wheeler and R.~P. Feynman, ``Interaction with the absorber as the mechanism of radiation,'' Rev. Mod. Phys. \textbf{17}, 157--181 (1945).
\item R.~P. Feynman, ``QED: The strange theory of light and matter,'' Princeton University Press, 1985, ISBN 0-691-08388-6.
\end{enumerate}

\section*{Author introduction:}
Vasant Natarajan heads the laser laboratory in the Department of Physics at the Indian Institute of Science. His research interests are in laser cooling of atoms, and using ultracold atoms to test time-reversal symmetry in the fundamental laws
of physics.

\end{document}